\documentclass[useAMS,usenatbib]{mn2e}
\usepackage{graphicx}
\usepackage{natbib}
\usepackage{aas_journals}
\usepackage{color}
\usepackage{amsmath}
\usepackage{times}

\newcommand{\ltsima}{$\; \buildrel < \over \sim \;$}
\newcommand{\lsim}{\lower.5ex\hbox{\ltsima}}
\newcommand{\gtsima}{$\; \buildrel > \over \sim \;$}
\newcommand{\gsim}{\lower.5ex\hbox{\gtsima}}

\newcommand{\avg}[1]{\langle{#1}\rangle}

\title[On the Coherence of WMAP and Planck Temperature Maps]{On the Coherence of WMAP and Planck Temperature Maps}
\author[A. Kov\'acs,  J. Carron and I. Szapudi]{Andr\'as Kov\'acs$^{1,2}$, 
Julien Carron$^{3}$ and Istv\'an Szapudi$^{3}$\\
$^{1}$ Institute of Physics, E\"otv\"os Lor\'and University, 1117 P\'azm\'any P\'eter s\'et\'any 1/A Budapest, Hungary\\
$^{2}$ MTA-ELTE EIRSA "Lendulet" Astrophysics Research Group\\
$^{3}$ Institute for Astronomy, University of Hawaii 2680 Woodlawn Drive, Honolulu, HI, 96822, USA}

\begin{document}

\date{Submitted 2013}

\pagerange{\pageref{firstpage}--\pageref{lastpage}} \pubyear{2013}

\maketitle

\label{firstpage}
\begin{abstract}
The recent data release of ESA's Planck mission together with
earlier WMAP releases provide the first opportunity to compare high resolution full sky Cosmic Microwave Background temperature anisotropy maps. 
To quantify the coherence of these maps beyond the power spectrum we introduce
Generalized Phases in the sense of SO(3), unit vectors in the $2\ell+1$ dimensional 
representation spaces.
For an isotropic Gaussian distribution, Generalized Phases point to random directions and if there is non-Gaussianity, they represent most of the non-Gaussian
information. The alignment of these unit vectors from two
maps can be characterized by
their angle, $0^\circ$ expected for full coherence, and $90^\circ$ for random vectors. We analyze maps from both missions
with the same mask and $N_{side}=512$ resolution, and
compare both power spectra and Generalized Phases. 
We find excellent agreement of the Generalized Phases of
Planck Smica map with that of the WMAP Q,V,W maps, 
rejecting the null hypothesis of
no correlations at $5\sigma$ for $\ell$'s $\ell<700$, $\ell<900$ 
and $\ell<1100$, respectively, except perhaps for $\ell<10$.
Using foreground reduced maps for WMAP increases the 
phase coherence.
The observed coherence angles can be explained
with a simple assumption of Gaussianity and a WMAP noise model
neglecting Planck noise, except for low-intermediate $\ell$'s there is a slight,
but significant off-set, depending on WMAP band. On the same scales WMAP
power spectrum is about $2.6\%$ higher 
at a very high significance, while at higher $\ell$'s there appears to be
no significant bias. Using our theoretical tools,
we predict the phase alignment of Planck with a hypothetical perfect
noiseless CMB experiment, finding decoherence at $\ell\simeq 2900$;
below this value Planck can be used most efficiently to constrain 
non-Gaussianity.

\end{abstract}


\section{Introduction}

One of the principal goals of modern cosmology is to characterize the statistical properties of the primordial density fluctuations, i.e. the seeds of the present large-scale structure. As widely presumed, the initial perturbations are associated with quantum properties of an inflationary field \citep{guth1981}. If this model is correct, the primordial fluctuations should be overwhelmingly Gaussian \citep{bardeen,bond} along with the small temperature
fluctuations of the Cosmic Microwave Background (CMB) sky.

Gaussianity is the most fundamental prediction of inflation.
Randomness of the complex phases of the harmonic coefficients
of small CMB temperature fluctuations provides natural constraints, since 
departures from Gaussian behavior typically cause deviations from
randomness \citep{colesNat}. There are several methods
constraining non-Gaussianity from phase information: phase mapping and uniformity tests \citep{chiang2002,chiang2004}, Shannon entropy of phases \citep{chiang2000}, surrogates \citep{raeth2010}, random walks \citep{stannard2005,hansen2011}, etc. These have been applied to WMAP all-sky maps, and most
recently to Planck \citep{planck_isotropy}. In some cases, non-Gaussian residuals have been detected \citep{chiang2003,naselsky2005},
although no primordial non-Gaussianity has been found with any
certainty.

Other studies, such as \cite{land2005,land2005b,land2006,copi2004b,copi2006} and \cite{bielewicz2005} defined directions on a sphere at each $\ell$ to construct estimators constraining unusual alignments and correlations in the harmonic series representing the CMB maps. Several ``anomalies'' and alignments were
identified, and several tests have been performed to explore their origin 
\citep{francis2010,frommert2010,rassat2013}.
These marginally significant anomalies were originally detected in WMAP, 
and recently confirmed in Planck \citep{planck_isotropy}.

Complex phases correspond to a unit vector in the complex plane, where the U(1) group acts as a rotation.
Based on this observation we generalize the usual U(1) phases for the group SO(3), relevant to the CMB or any full-sky map, as unit vectors in 
$(2\ell+1)$ dimensional representation spaces. These Generalized Phases in the sense of SO(3) respond to SO(3) rotations analogously to complex phases responding to U(1) rotations.
In the rest of this paper we only deal with the SO(3) group, therefore without ambiguity we can call them Generalized Phases, or GPs, hereafter.
For an isotropic Gaussian field, they correspond to a random direction by symmetry, represent
most of the information beyond the measured power spectrum, and they are independent from it. Nevertheless, two observations
of the same CMB realization should have exactly the same phases.
The principal aim of this work is to use this simple property
to construct a rigorous and concise $\ell$-by$\ell$ comparison of WMAP and Planck maps that emphasizes information beyond the
power spectrum. In particular, we will characterize coherence of two maps by the
angle of the unit vectors corresponding to their GPs, that also 
corresponds to a correlation coefficient in harmonic space.

We organize this paper as follows. In Section 2 we describe the data we used, and introduce our methods including theoretical expectations, simulations and measurements. In Section 3 our results, and statistical significances of our findings are presented. Finally, we briefly summarize our results in Section 4. 
The appendix contains derivations of formulae used in the main text.

\section{Data and methods}

To quantify the coherence of WMAP and Planck we first prepare maps
of the same resolution.
The WMAP team provides $N_{side}=512$ CMB temperature maps, therefore we choose this as our base resolution.
The Planck CMB products have higher resolution, $N_{side}=2048$, thus  
we downgraded Planck maps using HEALPIX \citep{healpix} for $N_{side}=512$.
We also used the $N_{side}=512$ WMAP9 Temperature Analysis Mask that
leaves 78\% of the sky for our analysis.

For WMAP, we used the Q,V,W frequency bands downloaded from the LAMBDA website \footnote{\texttt{http://lambda.gsfc.nasa.gov/}}, using both original and foreground reduced versions \citep{wmap,bennett2012}. For Planck, we downloaded the
NILC and Smica CMB maps \citep{pla2013} from the Planck Legacy Archive \footnote{\texttt{http://www.sciops.esa.int/index.php?project=PLANCK}}. They
already have galactic foregrounds and known point sources removed.

\subsection{Generalized Phases}

The CMB temperature fluctuations can be expanded into
spherical harmonics:
\begin{equation}
\frac{\Delta T}{\bar T} (\vartheta,\varphi) = \sum_{\ell=0}^{\infty} \sum_{m=-\ell}^{\ell} a_{\ell m} Y_{\ell m}(\vartheta,\varphi)
\label{eq_dT}
\end{equation}
Phases are defined by complex $a_{\ell m}$ coefficients of CMB multipoles as follows
\begin{equation}
a_{\ell m} = |a_{\ell m}| \cdot \exp(i\phi_{\ell m})
\label{eq35}
\end{equation}
These Fourier phases generate rotations around the $z$-axis,
corresponding to to the U(1) subgroup of the full SO(3) symmetry of the
harmonic coefficients. For Gaussian random fields (GRF), 
Fourier phases are random and uniformly distributed between 0 and $2\pi$. 
Testing the randomness of these phases therefore provides an 
interesting diagnostic of the Gaussianity of the fluctuation field \citep{colesNat}. Note that the power at each $\ell$ and these phases do not fully determine the random field.

To generalize complex Fourier phases, we first build 
$(2\ell+1)$ dimensional vectors using real and imaginary parts of $a_{lm}$ coefficients:
\begin{equation}
\varepsilon_{\ell}=(a_{\ell 0}/\sqrt{2}, Re[a_{\ell 1}], .... Re[a_{\ell \ell}], Im[a_{\ell 1}], .... Im[a_{\ell \ell}])
\label{xl}
\end{equation}
These vectors contain all the information due to the reality of the underlying
random field. For a Gaussian field, this is a random vector, with
each elements of $\varepsilon_{\ell}$ having a variance of $C_\ell/2$.
Generalized Phases are now defined as $(2\ell+1)$ dimensional unit vectors
\begin{equation}
\hat{\varepsilon}_{\ell} = \frac{\varepsilon_{\ell}}{\sqrt{\sum_{k}{ {\varepsilon^{2}_{\ell,k}} }}}.
\label{xl2}
\end{equation}
As $a_{\ell m}$ coefficients of different multipoles are independent, GPs are uncorrelated for a Gaussian distribution. Moreover, they follow uniform distributions over the sphere $S^{2\ell}$ for each $\ell$ \citep{cai2013}.
The statistics of GPs contain information complementary to the 
power spectrum, and for mildly non-Gaussian distributions, they 
should contain most of the non-Gaussian information. If the power
and the GPs are given, the realization of a random field is fully constrained.

In this work, we compare Generalized Phases to quantify the (generalized)
phase coherence of WMAP and Planck maps, i.e. the $\ell$-by$\ell$
coherence of the maps beyond and independently of the match of their
power spectra.

To quantify the coherence of the two maps, we calculated dot products of unit vectors defined by individual datasets at each $\ell$ multipole as
\begin{equation}
\cos\Theta_\ell = \sum_{k}\hat{\varepsilon}^{\rm Planck}_{\ell,k} \cdot \hat{\varepsilon}^{\rm WMAP}_{\ell,k}. 
\label{scal}
\end{equation}
\begin{figure}
\begin{center}
\includegraphics[width=90mm]{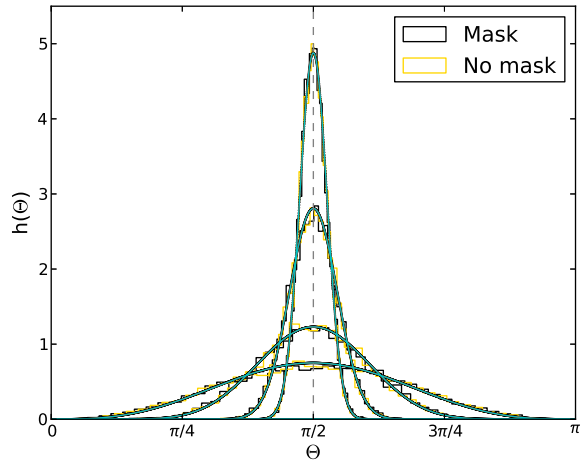}
\caption{Distributions of angles between random unit vectors in $(2\ell+1)$ dimensions. We compare analytic expectations and simulations, and test properties of galactic masks, as well. We present $\ell=2,5,25,75$ cases. These illustrate that angles are concentrated around $\pi /2$. This concentration becomes stronger as the dimension $n$ grows. These results are insensitive to the galactic
mask as long as the unit vectors are truly random.}
\label{sim}
\end{center}
\end{figure}
\subsection{Random angle statistics in $n$ dimensions}

Angles between Generalized Phases of two uncorrelated datasets - e.g.  CMB realizations - fluctuate around  $\pi/2$, their distributions are characterized by analytic formulae \citep{cai2013}. When the dimension $n=2\ell+1$ is fixed, 
the distribution of angles has a density function given by
\begin{equation} 
h(\Theta) = \frac{1}{\sqrt{\pi}} \frac{\Gamma(\frac{n}{2})}{\Gamma(\frac{n-1}{2})} \cdot \sin^{n-2} \Theta.
\label{htheta}
\end{equation}
Note that if $n$ = 2, $h(\Theta$) is the uniform density on [0,$\pi$]. When $n$ $\geq$ 3, h($\Theta$) is a unimodal  distribution with peak position of $\Theta$ = $\pi/2$. The concentration around $\pi/2$ becomes stronger as $n$ grows, since $\sin^{n-2}\Theta$ is driven to zero quickly for $\Theta \neq \pi /2$ \citep{cai2013}. This means that uncorrelated vectors in high dimensions tend to be perpendicular. As expected, in large dimensions, the distribution tends to a Gaussian distribution centered on  $\pi/2$. 
In Fig. \ref{sim} we show estimates of distributions of angles between unit vectors in higher dimensions. We simulated 500 CMB skies to test Eq. \eqref{htheta}. Simulations were made by using WMAP9 cosmological parameters, and WMAP9 noise.
We randomly choose 10,000 pairs of CMB simulations, and calculate Generalized Phases. Four examples of $\ell=2,5,25,75$ illustrate that individual distributions of angles between random unit vectors in $(2\ell+1)$ dimensions follow Eq. \eqref{htheta} closely. We checked that these results hold up to $\ell=1535$, the
maximum we can measure with our maps.

\begin{figure}
\begin{center}
\includegraphics[width=80mm]{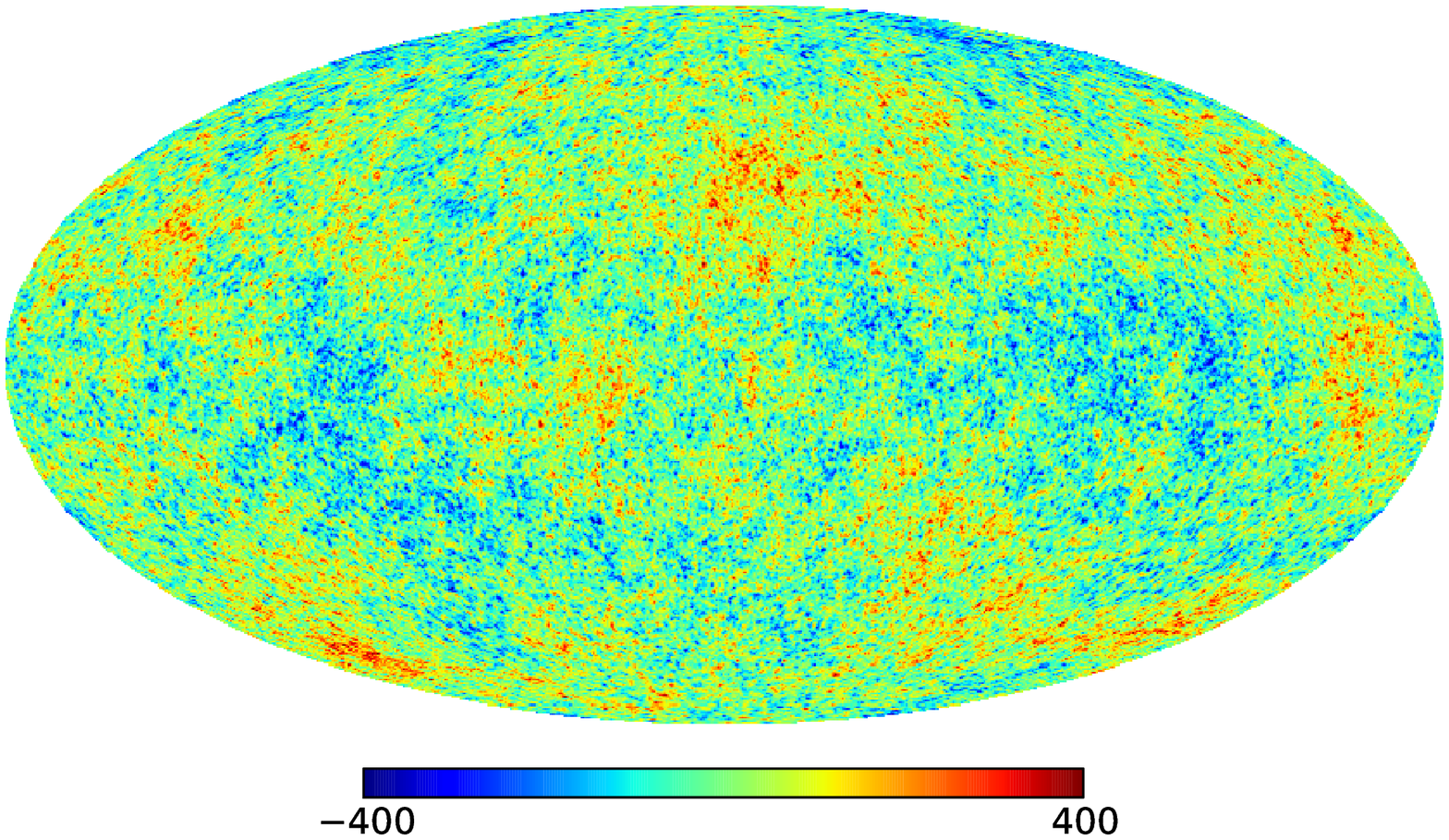}
\includegraphics[width=80mm]{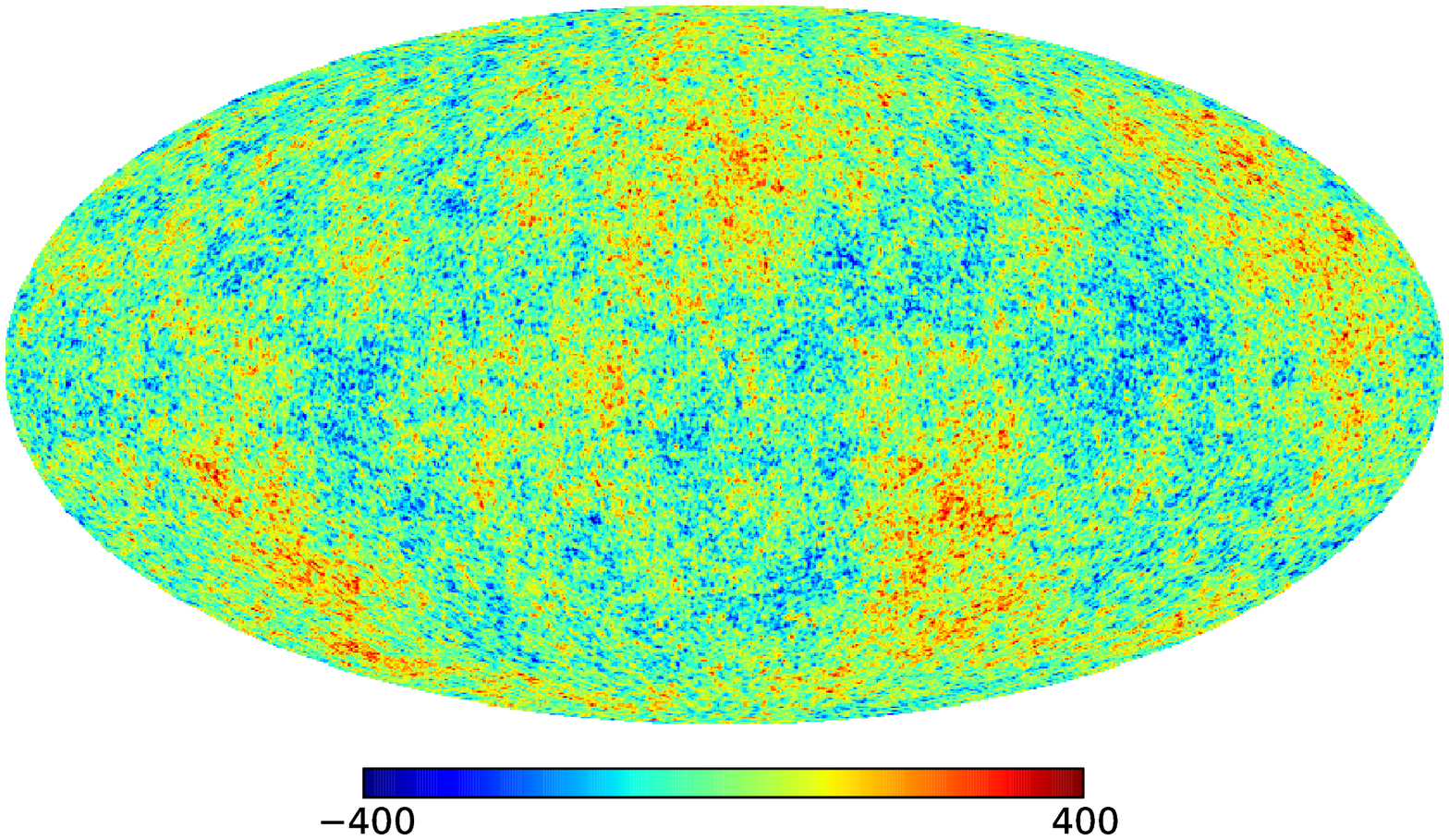}
\caption{We show original (amplitude-shuffled) version of a simulated CMB map on the top (bottom). These CMB maps have the same phases and pseudo power spectra, but different GPs.}
\label{GPs}
\end{center}
\end{figure}

We repeated our measurements on masked CMB skies using WMAP9 Temperature Analysis Mask. According to Fig. \ref{sim}, and perhaps somewhat surprisingly,
no difference was found. While galactic mask strongly affects statistical analysis of normal phases \citep{chiang2007}, the distribution of $\Theta$ is insensitive to the mask. The CMB mask is centered on $\vartheta=\pi/2$ in the spherical coordinate, which causes strong phase correlation only among phases of $\ell \approx m$.

It is useful to consider the closely related
correlation coefficient $\cos \Theta = C_\ell^{AB}/\sqrt{C_\ell^A C_\ell^B}$ in 
addition to $\Theta$. In this case, $x = (\cos \Theta + 1)/2$ follows the Beta distribution on $[0,1]$, i.e. $ x^{\alpha -1}(1-x)^{\beta - 1} / B(\alpha,\beta)$ with parameters $\alpha = \beta = (n-1)/2 = \ell$. Thus the exact first two moments of $\cos \Theta$ are
\begin{equation}
\left \langle {\cos \Theta}\right \rangle  = 0,\quad \left\langle\cos^2\Theta\right \rangle = \frac{1}{n}
\end{equation}

\newcommand{\CMB}{\textrm{CMB}}

\begin{figure}
\begin{center}
\includegraphics[width=90mm]{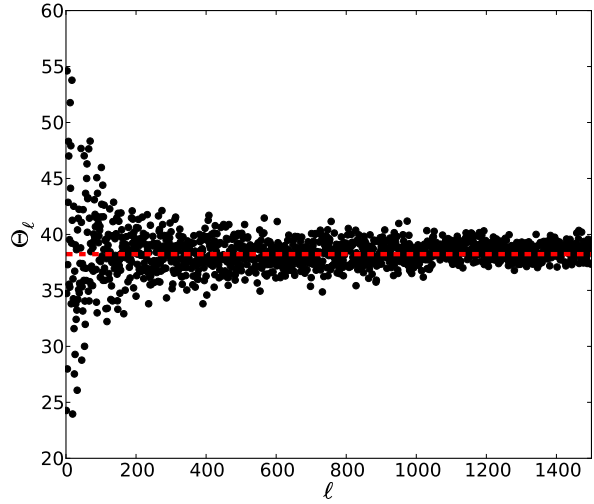}
\caption{Measured $\Theta_\ell$ angles of GPs of the original and $|a_{\ell m}|$-shuffled maps are illustrated. See text for details. }
\label{GPs2}
\end{center}
\end{figure}

Finally, we quantified the resolving power of GPs by the following procedure. We shuffled $|a_{\ell m}|$ amplitudes of a simulation for a given $\ell$, keeping both pseudo power spectrum and phases unchanged. Fig. \ref{GPs} shows the original and the "shuffled" CMB maps. We measured $\Theta_\ell$ angles between GPs of the maps (Fig. \ref{GPs2}), finding values fluctuating around $\Theta_\ell \approx 38^{\circ}$. We  integrated the Gaussian distributions of the $a_{\ell m}$'s to find the
average value $\avg{\cos \Theta_\ell} = \pi / 4$. This corresponds to $\Theta_\ell = 38.24^{\circ}$, i.e. $78.5\%$ correlation. 

\subsection{CMB and noise}

WMAP and Planck measurements of the CMB sky contain noise. This noise induces a rotation of the unit vectors $\hat{\varepsilon}^{\rm CMB}_{l}$ on the $2\ell$ dimensional sphere. Assuming full sky coverage and isotropic Gaussian noise, these rotations will only depend on the respective spectra of the $\CMB$ and that of the noise. The angles obey
\newcommand{\beq}{\begin{equation}}
\newcommand{\enq}{\end{equation}}
\newcommand{\noise}{\textrm{noise}}
\newcommand{\lp}{\left (}
\newcommand{\rp}{\right )}
\beq
\cos \Theta_\ell = \frac{\epsilon_\ell^{\CMB}  \cdot  \lp \epsilon_\ell^{\CMB}  +  \epsilon_\ell^{\noise} \rp}{\left | \epsilon_\ell^{\CMB} \right |  \left| \epsilon_\ell^{\CMB}+ \epsilon_\ell^{\noise} \right | }.
\enq
In the case of Gaussian noise, it is possible to obtain an explicit form for the distribution of the angle, generalizing \eqref{htheta}. Introducing the signal to noise $SN$ as the ratio of the norms of the two vectors,
\beq
SN = \frac{\left | \epsilon_\ell^\CMB \right |  }{ \left| \epsilon_\ell^\noise \right | } = \sqrt \frac{C_\ell^\CMB}{C_\ell^\noise},
\enq
one finds (see Appendix~\ref{appendixa} for details)
\beq\label{pdfsn}
\begin{split} 
&h_N\lp \Theta \rp = \frac{\Gamma(n)}{\Gamma\lp \frac{n-1}{2}\rp} \sin^{n-2} \Theta \\
& \cdot  \exp\lp-\frac n2 SN^2 \sin^2\Theta \rp 
\textrm{i}^{n-1} \textrm{erfc}\lp- \sqrt{\frac n 2 }  SN\cos \Theta\rp,
\end{split}
\enq
where the special functions
\beq
\textrm{i}^{n} \textrm{erfc}\lp z\rp = \frac 2 {\sqrt \pi} \int_z^\infty dt\: \frac{\lp t-z \rp^n}{n!} e^{-t^2}
\enq
are the iterated integrals of the complementary error function \citep{abramowitz70a}. These functions satisfy convenient recursion relations allowing easy generation of $h_N(\Theta)$. With the help of $\textrm{i}^{n} \textrm{erfc}\lp 0 \rp = 2^{-n}/\Gamma(n/2 + 1)$ we can check that we recover the corresponding distribution \eqref{htheta} for $\Theta$ in the limit of vanishing signal to noise, as expected. 

\begin{figure}
\begin{center}
\includegraphics[width=90mm]{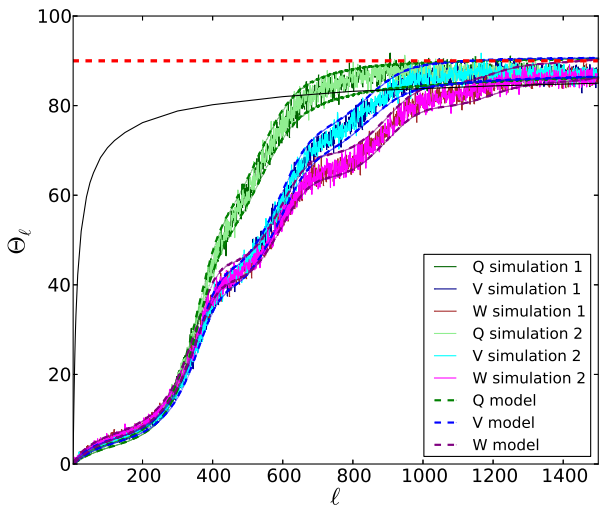}
\caption{We measured $\Theta_{\ell}$ angles between a Gaussian simulation, and the same simulation with WMAP noise added. We show two noise realizations of WMAP's Q, V, and W measurements, and compare them with 2$\sigma$ limits of our noise model. Dashed red line illustrates the expected value of $\pi / 2$ for no correlation, while the solid black curve shows 5$\sigma$ difference from this at each $\ell$.}
\label{noises}
\end{center}
\end{figure}

Again, the density function is very close to a Gaussian. Useful simple approximations for its mean and variance are
\beq
\left \langle{\cos \Theta} \right \rangle \approx \frac{SN}{\sqrt{1 +  SN^2}} = \sqrt{ \frac{C^{\CMB}_l}{C_l^{\CMB} + C_l^{\noise}}}
\label{mean}
\enq
and
\beq
\left \langle{\cos^2 \Theta} \right \rangle - \left\langle \cos \Theta \right \rangle^2 \approx  \frac{1}{n} \frac{1 + SN^2/2}{\lp 1 + SN^2 \rp^3}.
\label{var}
\enq
Both of these approximations are already at least $5\%$ accurate for any value of SN at $\ell = 5$. 
We evaluate Eqs. \eqref{mean} and \eqref{var} using WMAP Q, V, and W noise realizations, that are white noise to a good approximation, and represent different variances. We compared our model with simulations on Fig. \ref{noises}, and found that higher variance causes decoherence at lower $\ell$. Besides, different realizations of WMAP noise produced almost identical curves, in agreement with our model.

\section{Results}

\begin{figure}
\begin{center}
\includegraphics[width=90mm]{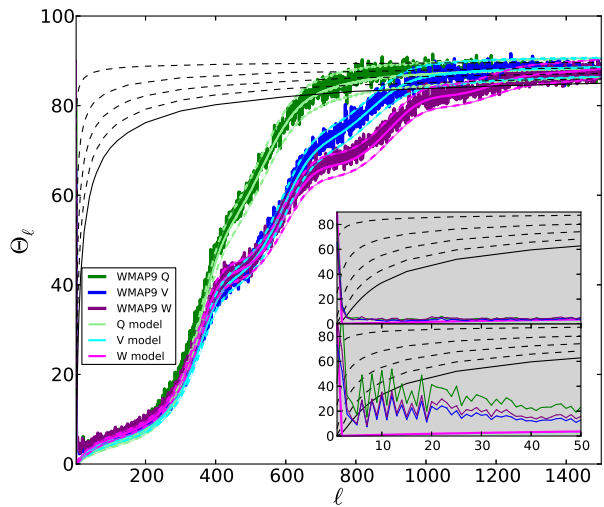}
\caption{The measured and theoretically predicted angles between Generalized Phases of Planck Smica map, and WMAP products. A black solid line defines the 5$\sigma$ alignment confidence level, while dashed black curves correspond to 4$\sigma$, 3$\sigma$, 2$\sigma$, and 1$\sigma$ values. The top inset zooms on $\ell<50$, while bottom inset shows the same without foreground cleaning.}
\label{res1}
\end{center}
\end{figure}

We obtained Generalized Phases of WMAP and Planck datasets by applying Equations \eqref{xl} and \eqref{xl2}. We present our results for the Planck Smica map, 
but repeating all our analysis with the NILC map produced
virtually identical results.
We used Eq. \eqref{scal} to characterize the coherence of the maps. 
While this angle does not contain all information, indeed there are many
ways of constructing a unit vector that is at angle $\Theta$ with respect
to another one, it corresponds to a concise way of expressing coherence,
and we can additionally interpret $\cos\Theta_{\ell}$ in terms of $C_\ell$'s is a
of correlation coefficient  $C^{\rm WMAP, Planck}_\ell/\sqrt{C_\ell^{\rm WMAP}C_\ell^{\rm Planck}}$, 
i.e. $60^\circ$  means 50\% correlation between the two maps.

To quantify the coherence, we choose as our null hypothesis that the
two maps are {\em not} correlated. In that case the distribution
$\Theta_\ell$ follows analytic distributions of Eq. \eqref{htheta}, and $p$-values can be calculated by integrating Eq. \eqref{htheta} to the measured $\Theta_\ell$. We define the two maps
as significantly correlated if the null hypothesis can be rejected at the
$5\sigma$ level.

\begin{figure}
\begin{center}
\includegraphics[width=80mm]{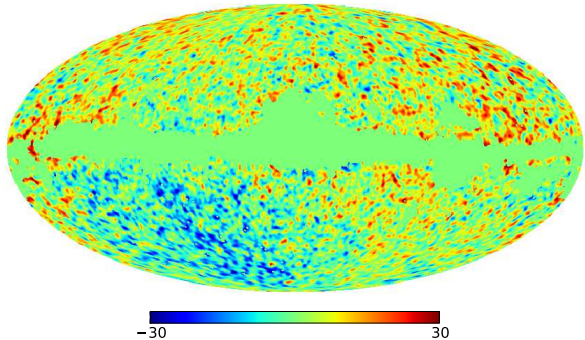}
\caption{Difference map of Planck Smica and WMAP9 Internal Linear Combination (ILC) maps in $\mu K$, smoothed at $2\deg$. See text for details.}
\label{diff}
\end{center}
\end{figure}

Figure \ref{res1} shows our results, where we compare Planck Smica map to WMAP Q, V and W band measurements. In general, the correlation between the maps
decreases with $\ell$, as qualitatively expected in the presence of uncorrelated
noise. For the lowest $\ell$'s the null hypothesis cannot be rejected at the
$5\sigma$ level, especially for the Q band, but using foreground reduced maps
improves the correlation to the point that maybe only the dipole 
is incoherent. This, however, only reflects the different cleaning procedures used by WMAP and Planck. In particular, the Smica algorithm sets $\ell=0,1$ exactly to zero, therefore it contains no information on the CMB (Jean-Francois Cardoso, private communication). The pattern illustrated on Figure \ref{diff} was also detected by  \cite{Fre2013}.

For higher $\ell$'s, the monotonically increasing $p$-values reach the limit confidence levels corresponding to $5\sigma$. We define these $\ell$'s corresponding to decoherence at $\ell\approx700$, $\ell\approx900$ and $\ell\approx1100$ for Q, V and W maps, respectively. This result is robust whether we use foreground removed WMAP maps or not, or Planck Smica/NILC maps. The observed decoherence can be fully explained based on a WMAP noise model, as illustrated
in our Figure~\ref{res1}, and explained in more detail next. Our interpretation is that WMAP GPs are dominated by noise above these $\ell$'s.

\begin{figure}
\begin{center}
\includegraphics[width=90mm]{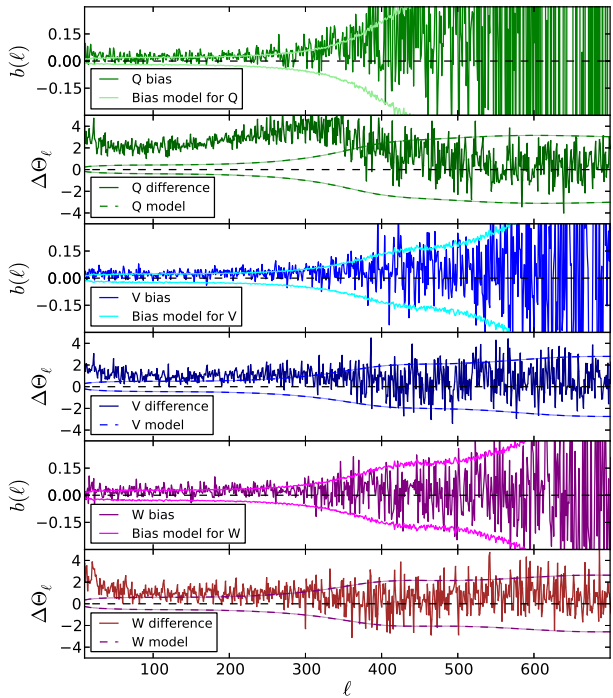}
\caption{We show measured biases of power spectra for Q, V and W bands, while estimated $2\sigma$ deviations are shown by solid lines. In addition, discrepancies between modeled and measured $\Theta_{\ell}$ are illustrated for Q, V and W, where dashed lines correspond to $2\sigma$ differences in our model.}
\label{bias}
\end{center}
\end{figure}

Our theory of Eq.~\ref{pdfsn} using simple Gaussian assumption for
both the CMB and noise provides a
prediction for  the expected coherence angle between the 
maps. The agreement is excellent with both
simulations and measurements at all $\ell$'s,
although there appears to be small but significant bias
in the measurements at low-intermediate $\ell$'s.
Figure~\ref{bias} displays the residual $\Theta_\ell$, i.e.
the difference between our theoretical predictions for the decoherence
based on our noise model, and the measured angle. For each Q,V and W,
there appears to be an excess angle, i.e. more decoherence than predicted,
for $\ell \lsim 500,400$ and
$300$, respectively. 

At face value in the framework
of our simple assumptions, this would be a sign of 
excess noise not taken into account in our noise model. It needs to
be emphasized though that this is a small, (although) significant 
effect, and therefore should be interpreted cautiously, given the assumption of uncorrelated 
Gaussian noise; noise correlations, foregrounds, and/or leakage from
the dipole (e.g. \cite{PrunetEtal2005,DasSouradeep}) could all influence the 
coherence angle in subtle ways.

For completeness, we measured power spectrum of the Planck Smica map, cross-power spectra of WMAP9 Q1-Q2, V1-V2, and an average cross-spectrum of six combinations of W1-W4 differential assemblies with SpICE \citep{SzapudiB}: the power spectrum is complementary to the GPs, corresponds to the
amplitude of the vector we defined in Eq.~\eqref{xl}, and might give
additional insight into the decoherence at low-intermediate $\ell$'s. 
We used WMAP9 beam transfer function products for 
Q1, Q2, V1, V2, W1, W2, W3, and W4 maps, 
and a 5' Gaussian smoothing for the Smica map.
We emphasize that we used again the same resolution maps, with the same mask,
and the same method to measure the power spectrum for all maps, thus our
comparison is more immediate than taking final products from the WMAP
and Planck team, respectively.

The power spectra are consistent with each other for the most part,
but curiously, in approximately the same range of $\ell$'s, where we found
less coherence than predicted by our theory, we find that $\tilde C_{\ell}^{\rm WMAP}$ is on average $2.6\%$ higher than $\tilde C_{\ell}^{\rm Planck}$ in the three Q,V,W maps. 
For the sake of consistency, we consider multipoles between $10$ and $300$
for each band, and find that 
the WMAP spectra are $2.7\%,2.6\% $ and $2.5\%$ higher than Smica, respectively. While visual inspection confirms the significance of this bias, we estimated it quantitatively in Appendix~\ref{appendixb} to be in the range of $10$'s of
$\sigma$'s. This bias is confined to these scales, the inclusion of higher multipoles result in a non-detection of significant bias.
While it would be difficult to assess quantitatively whether the bias
persists on larger $\ell$'s, at least qualitatively, it appears from
Figure~\ref{bias} that the bias is not significant above the
the same $\ell \gsim 500,400$ and $300$ for Q,V, and W, respectively,
where our theory predicts the decoherence based on the simple
Gaussian WMAP noise model. This might be a tantalizing hint, but
more investigations are needed to establish whether the two small,
but significant effects are related.

We repeated our measurements with WMAP 7 year foreground cleaned data, and found similar trends in terms of $\Theta_\ell$ angles. The agreement with WMAP9 results is less accurate, when we analyze maps without cleaning of foregrounds, but the difference is only significant at low $\ell$'s.
The most important observation, however, is that the estimated 
$5\sigma$ decoherence is at 
slightly lower $\ell$ if we use WMAP 7 year products. This is consistent 
with WMAP7 having more noise than WMAP9 further supporting the thesis
that all experiments observe the same underlying CMB, and that instrumental
noise causes the observed decoherence.

\newcommand{\masked}{\textrm{masked}}
\newcommand{\unmasked}{\textrm{unmasked}}
\newcommand{\fsky}{f_{\textrm{sky}}}

\subsection{Decoherence from WMAP noise and impact of mask at high $\ell$}

So far we established that the decoherence observed on Figure \ref{res1} is expected to originate primarily from the noise in WMAP. Assuming that the level of noise in the Smica map is negligible with respect to that of the Q,V,W maps on these scales, we can test this hypothesis using our density functions in Eq.~\eqref{pdfsn}. We proceeded as follows. Assuming white noise $\sigma^2_N$ in each WMAP maps, the signal to noise is $SN = C_{\ell}^\CMB / \sigma^2_N$, where $C_{\ell}^\CMB$ was generated with the CAMB package \footnote{\texttt{http://camb.info/}} with Planck's best fit parameters \citep{Planck_15}, multiplied by the respective beam window function of the Q,V or W  maps. The solid lines  in Figure \ref{res1} show the mean of the density function, and the dashed ones correspond to $2\sigma$ deviations. The decoherence is in excellent quantitative agreement with this simple model. It makes no difference to use the exact $h_N(\Theta)$ in  Eq.~\eqref{pdfsn} or the approximations in Eqs.~\eqref{mean}~and~\eqref{var}. 

The noise dominates by orders of magnitude at the highest $\ell$'s, therefore
an angle of $90$ degrees is expected naively. The observed angles, however,
deviate slightly from this theoretical prejudice, 
indicating a few percent residual correlation. As we show 
in more detail in Appendix~\ref{appendixc}, 
this correlation is due to leakage of low $\ell$ power 
into higher $\ell$'s, and essentially white noise.
We can obtain accurate analytic approximations assuming 
an azimuthally symmetric mask centered on the equator and white noise. The mask is an equatorial band sustaining an angle $b$ with the equator, so that $\fsky = 1-\sin b$. Using the explicit formula relating the spectrum $\tilde C_{\ell}$ of the masked field to that of the unmasked field $C_{\ell}$ \citep{2002ApJ...567....2H}, we derive in Appendix~\ref{appendixc} the asymptotic behavior of the spectrum,
\beq
\tilde C_\ell \rightarrow \quad \frac{16 \sigma^2_T \cos b}{\lp 2\ell +1 \rp^3},\quad \ell \rightarrow \infty
\enq
where $\sigma^2_T$ is the variance of the unmasked map
\beq
\sigma^2_T = \sum_{\ell} \frac{2\ell+1}{4\pi}C_{\ell} = \left \langle{ \lp \frac{\Delta T}{\bar T}\rp^2} \right\rangle. 
\enq
On the other hand the white noise spectra are simply multiplied by $\fsky$. Since $\left\langle \cos \Theta_l \right\rangle \approx \sqrt{\tilde C_l / (\tilde C_l + \sigma^2_N)}$, we obtain in the very low signal to noise regime
\beq
\begin{split}
\cos \Theta_{\ell}  \rightarrow   \frac{4}{\lp 2\ell + 1 \rp^{3/2}}  \lp \frac{\sigma_T}{\sigma_N} \rp \lp \frac{2-
\fsky}{\fsky} \rp^{1/4}
\end{split}
\enq
Despite the above approximations, these ideas explain the shape of measured $\Theta_{\ell}$ curves extremely well, and 
predict asymptotic properties at high $\ell$ in virtually perfect agreement with simulations and measurements. Note that these considerations do not affect
our 5$\sigma$ decoherence limits, as our null hypothesis of no correlations
(corresponding to infinite noise) has no bias.

\begin{figure}
\begin{center}
\includegraphics[width=90mm]{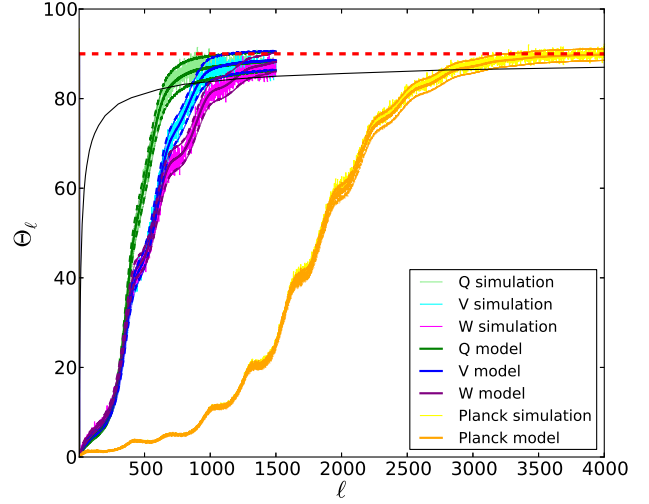}
\caption{Estimated decoherence properties of a Planck-like simulation ($N_{side}=2048$, assuming Smica noise) with a hypothetical perfect CMB experiment with no noise, and our WMAP results for comparison. We show simulated decoherence using single realizations, together with theory and $2\sigma$ errors. Black solid line illustrates $5\sigma$ significance level for our null hypothesis, while dashed red line shows $90^{\circ}$ decoherence level.}
\label{estim}
\end{center}
\end{figure}

We used our well calibrated decoherence model to forecast GP angles of Planck and a hypothetical perfect CMB experiment without noise (Fig. \ref{estim}). Decoherence is predicted at $\ell\approx 2900$, beyond which any non-Gaussian information should be dominated by noise.

\section{Conclusions}

We quantified the $\ell$-by-$\ell$ coherence of
latest  WMAP, and Planck CMB maps. We introduced a new set of statistics, Generalized Phases, that are complementary to the (pseudo-)power spectrum, and can
be used to characterize the phase-coherence of two CMB maps.
We compared GP's of the two maps by simply calculating the
angles between the corresponding unit vectors. These angles, while
do not contain all non-Gaussian information, concisely summarize the
coherence properties of two maps at each $\ell$.
Using the statistics of random vectors in $(2\ell+1)$ dimensions,
we defined the $\ell$ of decoherence where the null hypothesis of 
no correlation between the maps could not be rejected at the $5\sigma$
level. We controlled any effect of the masks, typically a problem
with statistics based on phases, with careful simulations and analytical
models that, albeit based on simplifying assumptions, appear to provide
an excellent quantitative framework. To check for systematics, we
repeated all our measurements of the Planck Smica map with the NILC maps
finding virtually identical results. According to our definition,
decoherence from Planck was found above $\ell\approx 700$, $\ell\approx 900$ and $\ell\approx 1100$ for WMAP9 Q, V and W.
Our theoretical description is in excellent agreement with the measured
coherence angles, with a slight bias for low-intermediate $\ell$'s
We also find
a small bias of the WMAP pseudo-$\tilde C_{\ell}$ at  $10 \le \ell \le 300$ 
at an average $2.6$\% level with very high significance.
It appears that for high $\ell$'s, where our theoretical prediction
for the coherence angle is accurate based on a simple Gaussian WMAP noise
model, there is no significant bias in the power spectra either.
Qualitatively, there is a slight color dependency as well based on
Figure~\ref{bias}. From the excess decoherence we can calculate the
amount of excess noise it corresponds to. We found that,
with the exception of the Q map in the range of 
$250 \lsim \ell \lsim  500$ the noise
corresponding to the excess decoherence is below what is needed
to fully explain the bias in the power spectra. Nevertheless,
the qualitative behaviour of the noise is similar to the observed
one, and it is different than our simulations.
In conclusion, there are tantalizing coincidences hinting
that the excess decoherence and power spectrum bias are related,
but no consistent picture emerged. Note that our simulations
do not contain correlated noise, we did not check for any effect
of foregrounds or low-$\ell$ leakage, especially from the dipole,
into higher $\ell$'s; such investigations 
are left for future research.

Our analytical and simulation framework can be used to
forecast the coherence of Planck with a noise-free experiment (the true
CMB). We find that below $\ell\le 2900$ Planck is coherent with the 
CMB according to our $5\sigma$ criterion, 
thus non-Gaussian information can be best gleaned from below these
$\ell$'s.

\section*{Acknowledgments}
We acknowledge support from NASA grants NNX12AF83G and NNX10AD53G. In addition, AK acknowledges support from Campus Hungary fellowship program, and OTKA through grant no. 101666.
\bibliographystyle{mn2e}
\bibliography{refs}

\begin{thebibliography}{}

\bibitem[\protect\citeauthoryear{Abramowitz \& Stegun}{Abramowitz \&
  Stegun}{1970}]{abramowitz70a}
Abramowitz M.,  Stegun I.,  1970, {Handbook of mathematical functions}.
{Dover Publications Inc.}, {New York}

\bibitem[\protect\citeauthoryear{{Bardeen}, {Bond}, {Kaiser} \&
  {Szalay}}{{Bardeen} et~al.}{1986}]{bardeen}
{Bardeen} J.~M.,  {Bond} J.~R.,  {Kaiser} N.,    {Szalay} A.~S.,  1986, \apj,
  304, 15

\bibitem[\protect\citeauthoryear{{Bennett}, {Larson} \& {Weiland}}{{Bennett}
  et~al.}{2012}]{bennett2012}
{Bennett} C.~L.,  {Larson} D.,    {Weiland} J.~L. e.~a.,  2012, ArXiv e-prints

\bibitem[\protect\citeauthoryear{{Bielewicz}, {Eriksen}, {Banday}, {G{\'o}rski}
  \& {Lilje}}{{Bielewicz} et~al.}{2005}]{bielewicz2005}
{Bielewicz} P.,  {Eriksen} H.~K.,  {Banday} A.~J.,  {G{\'o}rski} K.~M.,
  {Lilje} P.~B.,  2005, \apj, 635, 750

\bibitem[\protect\citeauthoryear{{Bond} \& {Efstathiou}}{{Bond} \&
  {Efstathiou}}{1987}]{bond}
{Bond} J.~R.,  {Efstathiou} G.,  1987, \mnras, 226, 655

\bibitem[\protect\citeauthoryear{{Cai}, {Fan} \& {Jiang}}{{Cai}
  et~al.}{2013}]{cai2013}
{Cai} T.,  {Fan} J.,    {Jiang} T.,  2013, Technical report of the Department
  of Statistics, University of Pennsylvania

\bibitem[\protect\citeauthoryear{{Chiang} \& {Coles}}{{Chiang} \&
  {Coles}}{2000}]{chiang2000}
{Chiang} L.-Y.,  {Coles} P.,  2000, \mnras, 311, 809

\bibitem[\protect\citeauthoryear{{Chiang}, {Coles} \& {Naselsky}}{{Chiang}
  et~al.}{2002}]{chiang2002}
{Chiang} L.-Y.,  {Coles} P.,    {Naselsky} P.,  2002, \mnras, 337, 488

\bibitem[\protect\citeauthoryear{{Chiang} \& {Naselsky}}{{Chiang} \&
  {Naselsky}}{2007}]{chiang2007}
{Chiang} L.-Y.,  {Naselsky} P.~D.,  2007, \mnras, 380, L71

\bibitem[\protect\citeauthoryear{{Chiang}, {Naselsky} \& {Coles}}{{Chiang}
  et~al.}{2004}]{chiang2004}
{Chiang} L.-Y.,  {Naselsky} P.~D.,    {Coles} P.,  2004, \apjl, 602, L1

\bibitem[\protect\citeauthoryear{{Chiang}, {Naselsky}, {Verkhodanov} \&
  {Way}}{{Chiang} et~al.}{2003}]{chiang2003}
{Chiang} L.-Y.,  {Naselsky} P.~D.,  {Verkhodanov} O.~V.,    {Way} M.~J.,  2003,
  \apjl, 590, L65

\bibitem[\protect\citeauthoryear{{Coles} \& {Chiang}}{{Coles} \&
  {Chiang}}{2000}]{colesNat}
{Coles} P.,  {Chiang} L.-Y.,  2000, \nat, 406, 376

\bibitem[\protect\citeauthoryear{{Copi}, {Huterer}, {Schwarz} \&
  {Starkman}}{{Copi} et~al.}{2006}]{copi2006}
{Copi} C.~J.,  {Huterer} D.,  {Schwarz} D.~J.,    {Starkman} G.~D.,  2006,
  \mnras, 367, 79

\bibitem[\protect\citeauthoryear{{Copi}, {Huterer} \& {Starkman}}{{Copi}
  et~al.}{2004}]{copi2004b}
{Copi} C.~J.,  {Huterer} D.,    {Starkman} G.~D.,  2004, \prd, 70, 043515

\bibitem[\protect\citeauthoryear{{Das} \& {Souradeep}}{{Das} \&
  {Souradeep}}{2013}]{DasSouradeep}
{Das} S.,  {Souradeep} T.,  2013, ArXiv e-prints

\bibitem[\protect\citeauthoryear{{Francis} \& {Peacock}}{{Francis} \&
  {Peacock}}{2010}]{francis2010}
{Francis} C.~L.,  {Peacock} J.~A.,  2010, \mnras, 406, 14

\bibitem[\protect\citeauthoryear{{Frejsel}, {Hansen} \& {Liu}}{{Frejsel}
  et~al.}{2013}]{Fre2013}
{Frejsel} A.,  {Hansen} M.,    {Liu} H.,  2013, JCAP, 6, 5

\bibitem[\protect\citeauthoryear{{Frommert} \& {En{\ss}lin}}{{Frommert} \&
  {En{\ss}lin}}{2010}]{frommert2010}
{Frommert} M.,  {En{\ss}lin} T.~A.,  2010, \mnras, 403, 1739

\bibitem[\protect\citeauthoryear{{Gorski}, {Hivon} \& et al.}{{Gorski}
  et~al.}{2005}]{healpix}
{Gorski} K.~M.,  {Hivon} E.,    et al. 2005, \apj, 622, 759

\bibitem[\protect\citeauthoryear{{Guth}}{{Guth}}{1981}]{guth1981}
{Guth} A.~H.,  1981, \prd, 23, 347

\bibitem[\protect\citeauthoryear{{Hansen}, {Frejsel}, {Kim}, {Naselsky} \&
  {Nesti}}{{Hansen} et~al.}{2011}]{hansen2011}
{Hansen} M.,  {Frejsel} A.~M.,  {Kim} J.,  {Naselsky} P.,    {Nesti} F.,  2011,
  \prd, 83, 103508

\bibitem[\protect\citeauthoryear{{Hivon}, {G{\'o}rski}, {Netterfield}, {Crill},
  {Prunet} \& {Hansen}}{{Hivon} et~al.}{2002}]{2002ApJ...567....2H}
{Hivon} E.,  {G{\'o}rski} K.~M.,  {Netterfield} C.~B.,  {Crill} B.~P.,
  {Prunet} S.,    {Hansen} F.,  2002, \apj, 567, 2

\bibitem[\protect\citeauthoryear{{Jarosik}, {Bennett} \& et al.}{{Jarosik}
  et~al.}{2011}]{wmap}
{Jarosik} N.,  {Bennett} C.~L.,    et al. 2011, \apjs, 192, 14

\bibitem[\protect\citeauthoryear{{Land} \& {Magueijo}}{{Land} \&
  {Magueijo}}{2005a}]{land2005}
{Land} K.,  {Magueijo} J.,  2005a, Physical Review Letters, 95, 071301

\bibitem[\protect\citeauthoryear{{Land} \& {Magueijo}}{{Land} \&
  {Magueijo}}{2005b}]{land2005b}
{Land} K.,  {Magueijo} J.,  2005b, \mnras, 362, 838

\bibitem[\protect\citeauthoryear{{Land} \& {Magueijo}}{{Land} \&
  {Magueijo}}{2007}]{land2006}
{Land} K.,  {Magueijo} J.,  2007, \mnras, 378, 153

\bibitem[\protect\citeauthoryear{{Naselsky}, {Chiang}, {Novikov} \&
  {Verkhodanov}}{{Naselsky} et~al.}{2005}]{naselsky2005}
{Naselsky} P.~D.,  {Chiang} L.-Y.,  {Novikov} I.~D.,    {Verkhodanov} O.~V.,
  2005, International Journal of Modern Physics D, 14, 1273

\bibitem[\protect\citeauthoryear{{Planck Collaboration}, {Ade}, {Aghanim},
  {Armitage-Caplan} \& et al.}{{Planck Collaboration}
  et~al.}{2013a}]{Planck_15}
{Planck Collaboration} {Ade} P.~A.~R.,  {Aghanim} N.,  {Armitage-Caplan}   et
  al. 2013a, ArXiv e-prints

\bibitem[\protect\citeauthoryear{{Planck Collaboration}, {Ade}, {Aghanim},
  {Armitage-Caplan} \& et al.}{{Planck Collaboration} et~al.}{2013b}]{pla2013}
{Planck Collaboration} {Ade} P.~A.~R.,  {Aghanim} N.,  {Armitage-Caplan} C.,
  et al. 2013b, ArXiv e-prints

\bibitem[\protect\citeauthoryear{{Planck Collaboration}, {Ade}, {Aghanim},
  {Armitage-Caplan} \& et al.}{{Planck Collaboration}
  et~al.}{2013c}]{planck_isotropy}
{Planck Collaboration} {Ade} P.~A.~R.,  {Aghanim} N.,  {Armitage-Caplan} C.,
  et al. 2013c, ArXiv e-prints

\bibitem[\protect\citeauthoryear{{Prunet}, {Uzan}, {Bernardeau} \&
  {Brunier}}{{Prunet} et~al.}{2005}]{PrunetEtal2005}
{Prunet} S.,  {Uzan} J.-P.,  {Bernardeau} F.,    {Brunier} T.,  2005, \prd, 71,
  083508

\bibitem[\protect\citeauthoryear{{Raeth}, {Rossmanith}, {Morfill}, {Banday} \&
  {Gorski}}{{Raeth} et~al.}{2010}]{raeth2010}
{Raeth} C.,  {Rossmanith} G.,  {Morfill} G.,  {Banday} A.~J.,    {Gorski}
  K.~M.,  2010, ArXiv e-prints

\bibitem[\protect\citeauthoryear{{Rassat}, {Starck} \& {Dupe}}{{Rassat}
  et~al.}{2013}]{rassat2013}
{Rassat} A.,  {Starck} J.-L.,    {Dupe} F.-X.,  2013, ArXiv e-prints

\bibitem[\protect\citeauthoryear{{Stannard} \& {Coles}}{{Stannard} \&
  {Coles}}{2005}]{stannard2005}
{Stannard} A.,  {Coles} P.,  2005, \mnras, 364, 929

\bibitem[\protect\citeauthoryear{{Szapudi}, {Prunet} \& {Colombi}}{{Szapudi}
  et~al.}{2001}]{SzapudiB}
{Szapudi} I.,  {Prunet} S.,    {Colombi} S.,  2001, \apjl, 561, L11

\end{thebibliography}
\begin{appendix}
\newcommand{\WMAP}{\textrm{WMAP}}
\newcommand{\alm}{a_{\ell m}^\CMB}
\newcommand{\alms}{a_{\ell m}^{\CMB,*}}
\newcommand{\av}[1]{\left \langle #1 \right \rangle}
\newcommand{\lb}{\left [}
\newcommand{\rb}{\right ]}
\renewcommand{\min}{\textrm{min}}
\renewcommand{\max}{\textrm{max}}

\section{The distribution of coherence angles for noisy data}
\label{appendixa}

We derive the form of Eq.~\eqref{pdfsn} next. The probability density for $\cos \Theta_\ell$ is given by
\beq \label{pdfcos}
\begin{split}
p(\cos \Theta) &= \int d^n \epsilon_\ell \:p_G \lp \epsilon_\ell \rp \\
&\cdot \delta^D\lp \cos \Theta_l - \hat \epsilon^\CMB_\ell \cdot \lp \hat \epsilon^\CMB_\ell + \hat \epsilon_\ell \rp \rp,
\end{split}
\enq
with $\delta^D$ the Dirac delta function and $p_G$ is the probability density describing $n$ Gaussian uncorrelated variables with variance $C_\ell^\noise/2$. We can set without loss of generality $\epsilon^\CMB$ to be parallel to the first axis, such that
\beq
 \hat \epsilon^\CMB_\ell \cdot \lp \hat \epsilon^\CMB_\ell + \hat \epsilon_\ell \rp = \frac{C_\ell^\CMB + \epsilon^1_\ell}{\sqrt{(C_\ell^\CMB + \epsilon^1_\ell)^2 + \sum_{k >1} \lp \epsilon^k_\ell \rp^2}}.
\enq
Shifting the variable $\epsilon^1_\ell \rightarrow C_\ell^\CMB + \epsilon^1_\ell$ in Eq. \eqref{pdfcos} we simplify the integral further. The argument of the integrand depends only of the radial coordinate and of the first polar angle defined by $\epsilon^1_\ell = r \cos \phi_1$, which must match $\Theta_\ell$, because of the Dirac delta function. In $n$-dimensional space we have
\beq
 d^nx = r^{n-1}dr \sin^{n-2} \phi_1 d\phi_1 \cdots.
\enq
The Dirac delta function gives the factor $\sin^{n-2}\Theta$ in Eq.~\eqref{pdfsn}, and the radial integral the second factor.

\section{Estimate of bias significance} 
\label{appendixb}
\renewcommand{\CMB}{\textrm{Planck}}
We define the bias of WMAP with respect to Planck at a given $\ell$ as
\beq
b_\ell = \frac{C^{\WMAP}_\ell}{C^{\CMB}_\ell} - 1.
\enq
We expect $C_\ell^\WMAP$ to coincide on average with $C^{\CMB}_\ell$ , in which case $\av{b_\ell} = 0$. Our aim is to estimate $\av{b^2_\ell}$. We need to make some
simplifying assumptions on the stochasticity of $C_\ell^\WMAP$. We assume that this stochasticity comes from the cross-correlation of two noisy tracers,
\beq
C_\ell^\WMAP =  \frac{1}{2\ell + 1}\sum_{m = -\ell}^\ell \lp \alm + \epsilon_{1,\ell m} \rp\lp \alms + \epsilon^*_{2,\ell m} \rp,
\enq
We assume that the harmonic coefficients $\epsilon_{\ell m}$ of the noise are Gaussian variables  with spectrum $\av{\epsilon_{i,\ell m}\epsilon^*_{j,\ell' m'}} = \delta_{\ell \ell'}\delta_{mm'}\delta_{ij} C^{N_i},\quad i,j = 1,2 $, while $\alm$ and $C^\CMB_\ell$ are simple numbers. Within these assumptions it holds that
\beq
b_\ell =  \frac{1}{ C_\ell^\CMB\lp 2\ell + 1 \rp} \sum_{m = -\ell}^\ell \lp \alm\epsilon_{2,\ell m}^* + \alms\epsilon_{1,\ell m} + \epsilon_{1,\ell m}\epsilon_{2,\ell m}^* \rp.
\enq
Averaging over noise gives no bias, and the variance of $b_\ell$ can be simply evaluated remembering that we treat $\alm$ as simple numbers and that the average of three $\epsilon$'s vanishes. We obtain
\beq
\av{ b_\ell^2} = \frac{1}{2\ell + 1}  \lb  \lp \frac{C_\ell^{N_1} + C_\ell^{N_2}}{C_\ell^\CMB}  \rp + \lp \frac{C_\ell^{N_1} C_\ell^{N_2}}{C_\ell^\CMB}  \rp^2 \rb.
\enq
Averaging over multipoles defines the bias $b = \lp \sum_{\ell}b_\ell \rp / \Delta\ell$. Neglecting correlations, $\av{b^2_\ell} =  \lp \sum_\ell \av{b_\ell}^2  \rp/ \lp \Delta \ell \rp^2$. We set further $C_\ell^{N_1} = C_\ell^{N_2} = 2C_\ell^N$, defining $C_\ell^\CMB / C^N_\ell$ as the signal to noise $SN_\ell$ of the map. Thus we obtain our final formula
\beq
\av{b^2} = \frac{4}{ \lp \Delta \ell \rp^2} \sum_{\ell_{\textrm{min}}}^{\ell_{\textrm{max}}} \lp  \frac{1 + SN_\ell}{ \lp 2\ell + 1\rp \ \lp SN_\ell \rp^2} \rp,
\enq
with $\Delta \ell = \ell_\max - \ell_\min + 1$, with which we estimated the significance of the bias. 

For a roughly constant signal to noise a simple estimate of $\sigma_b = \sqrt{\av{b^2}}$ is
\beq
\sigma_b \approx  \frac{2 \sqrt {1 +SN}}{\Delta \ell \:SN}  \sqrt{\ln \lp \frac{\ell_\max}{\ell_\min}\rp}.  
\enq
Using the above formula and neglecting correlations between $C_\ell$'s,
we estimate the significance of the bias in the Q,V,W colors to be $33\sigma, 30\sigma$ and $26\sigma$, respectively. While taking into account the true covariance matrix, potentially impacted by correlated noise and mask,
could lower these significances, it is safe to state that the
bias below $\ell \lsim 300$ is overwhelmingly significant. At the same time,
if $\ell$'s up to $1100$ - the maximum given by $\ell_{max}$ of Q1,Q2 beam transfer fuctions - are taken into account, we find $1.6\sigma,0.7\sigma$, and $1.2\sigma$, i.e. no significant bias is detected over the full range of the power spectrum. Note, however, that
this is mainly due to the noise dominating at high $\ell$ and the
uniform weighting of our estimator, that is suboptimal for the bias
once the noise is increasing due to the tail of the beam correction.
\renewcommand{\CMB}{\texrm{CMB}}

\section{Coherence angle asymptotics with azymuthally symmetric mask } 
\label{appendixc}
We derive the asymptotic behavior of the coherence angle in the presence of an
azymuthally symmetric mask (band).
Our starting point is the exact formula relating the spectrum of the original map to that of the masked map \citep{2002ApJ...567....2H}
\beq
C^{\masked}_\ell = \sum_{\ell_2\ell_3} =\frac{2\ell_2 + 1}{4\pi} C_{\ell_2} |W_{\ell_30}|^2 \begin{pmatrix} \ell & \ell_2 & \ell_3 \\ 0 & 0 & 0  \end{pmatrix}^2.
\enq
In this equation $W_{l0}$ are the harmonic coefficient of the azimuthally symmetric mask function. We are interested in the regime where $\ell \rightarrow \infty$. In this case, it is possible to rewrite the above equation as follows,
\beq C^{\masked}_\ell  \stackrel{\ell \rightarrow \infty}{\rightarrow}   \frac 1{2\ell +1} \sum_{\ell_2} \frac{2\ell_2 + 1}{4\pi} C_{\ell_2} \left\langle |W_{\ell0}^2| \right\rangle_{\ell-\ell_2,\ell+\ell_2},
\enq
where the last term is the average of $|W_{\ell 0}^2|$ with a roughly flat weight function centered on $l$ with width $l_2$. The exact weight function can be obtained from the asymptotics of the Wigner $3j$ symbols \citep[e.g.]{2002ApJ...567....2H}
 but they turn out irrelevant for our purpose.  For a band mask centered on the equator with angle $b$, and thus $\fsky = 1-\sin b$, we have
\beq
W_{l0} = \sqrt{\frac{4\pi}{2\ell +1}} \lp P_{\ell-1}(\sin b) - P_{\ell + 1}(\sin b) \rp, \quad \ell \textrm{  even}.
\enq
where $P_\ell(x)$ are the Legendre polynomials. The coefficients for $\ell $ odd vanish due to the symmetry with respect to the equator. The polynomials have the asymptotic behavior
\beq
P_\ell(\cos \theta) \rightarrow \frac{2 \cos \lp (\ell + 1/2)\theta -\frac{\pi}{4}\rp} {\sqrt{ \pi  (2\ell +1) \sin\theta} },
\enq
at high $\ell$. Using this formula and the addition formula for sines and cosines, one has after some algebra
\beq \label{W2}
W^2_{l0} \rightarrow \frac{64}{\lp 2\ell + 1\rp^2} \sin \theta \sin^2 \lp (\ell + 1/2)\theta - \frac \pi 4 \rp,\quad \ell \textrm{  even},
\enq
with $\theta = \pi/2 - b$ and $\sin \theta = \cos b$.
We need the mean value of \eqref{W2} with respect to a smooth function centered on $\ell$ with size $2\ell_2$, small with respect to $\ell$. We can replace the $\sin^2$ in \eqref{W2} by a factor $1/2$, as the average would be the same if $\sin^2$ was in fact a $\cos^2$. Another factor $1/2$ comes from the fact that only even $\ell$ are non-zero. 
\beq 
\begin{split}
C^{\masked}_\ell  &\stackrel{\ell \rightarrow \infty}{\rightarrow}   \frac 1{2\ell +1} \sum_{\ell_2} \frac{2\ell_2 + 1}{4\pi} C_{\ell_2} \left\langle |W_{\ell0}^2| \right\rangle_{\ell-\ell_2,\ell+\ell_2} \\
&\sim \frac{16 \cos b}{2\ell + 1} \sum_{\ell 2}\frac{\lp 2\ell_2 + 1\rp}{4\pi} C_{\ell_2}\av{\frac{1}{\lp 2\ell + 1\rp^2}}_{\ell -\ell_2, \ell + \ell_2 }
\end{split}
\enq
If  $C_{\ell_2}$ only for $\ell_2$ much smaller than $\ell$, then the mean value becomes independent of $\ell_2$. All in all, we obtain
\beq
C_\ell^\masked \rightarrow \frac{16 \cos b}{ \lp 2\ell + 1 \rp^3}\lp \sum_{\ell 2}  \frac{\lp 2\ell_2 + 1\rp}{4\pi}C_{\ell_2}  \rp
\enq
 The last term is the variance of the map. This formula is valid both for small or large $\fsky$.
\end{appendix}
\end{document}